\documentclass[
10pt, 
a4paper, 
oneside, 
headinclude,footinclude, 
BCOR5mm, 
]{scrartcl}

\usepackage[
nochapters, 
beramono, 
eulermath,
pdfspacing, 
dottedtoc 
]{classicthesis} 

\usepackage{hyperref}

\usepackage{arsclassica} 

\usepackage[T1]{fontenc} 

\usepackage[utf8]{inputenc} 

\usepackage{graphicx} 
\graphicspath{{Figures/}} 

\usepackage{enumitem} 

\usepackage{subfig} 

\usepackage{amsmath,amssymb,amsthm} 

\usepackage{varioref} 

\usepackage{accents}

\usepackage[hmarginratio=2:1,top=25mm,left=40mm,columnsep=25pt]{geometry}

\usepackage{doi}

\theoremstyle{definition} 

\theoremstyle{plain} 

\theoremstyle{remark} 

\hypersetup{
colorlinks=true, breaklinks=true, bookmarks=true,bookmarksnumbered,
urlcolor=webbrown, linkcolor=RoyalBlue, citecolor=webgreen, 
pdftitle={}, 
pdfauthor={\textcopyright}, 
pdfsubject={}, 
pdfkeywords={}, 
pdfcreator={pdfLaTeX}, 
pdfproducer={LaTeX with hyperref and ClassicThesis} 
}

  \usepackage[numbers]{natbib}

\usepackage{empheq}
\newlength\mytemplen
\newsavebox\mytempbox
\makeatletter
\definecolor{cream}{rgb}{.81, .88, 1}
 \newcommand\mycreambox{%
     \@ifnextchar[
        {\@mycreambox}%
        {\@mycreambox[0pt]}}
 \def\@mycreambox[#1]{%
     \@ifnextchar[
        {\@@mycreambox[#1]}%
        {\@@mycreambox[#1][0pt]}}
 \def\@@mycreambox[#1][#2]#3{
     \sbox\mytempbox{#3}%
     \mytemplen\ht\mytempbox
     \advance\mytemplen #1\relax
     \ht\mytempbox\mytemplen
     \mytemplen\dp\mytempbox
     \advance\mytemplen #2\relax
     \dp\mytempbox\mytemplen
     \colorbox{cream}{\hspace{1em}\usebox{\mytempbox}\hspace{1em}}}
 \makeatother 

\usepackage[dvipsnames]{xcolor} 
\usepackage{amsmath}
\usepackage[mathscr]{eucal}
\usepackage{bm}
\allowdisplaybreaks
\usepackage{comment}

\usepackage{relsize} 

\newcommand{\Lag}{\Lambda}
\newcommand{\Act}{\mathcal{S}}

\newcommand{\spacetime}{\mathcal{V}^4}
\newcommand{\mattertime}{\mathcal{M}^4}

\newcommand{\paperI}{{\cite{GRGPR}}}
\newcommand{\pot}{J}
\newcommand{\potw}{w}
\newcommand{\potexp}[1]{w^{[#1]}}
\newcommand{\Temp}{T}
\newcommand{\ETJ}{\bar{E}}
\newcommand{\ETW}{\check{E}}

\newcommand{\pd}{\partial}
\newcommand{\rmd}{{\textrm d}}
\newcommand{\cgrad}[2]{ \xi^{#1}_{\phantom{#1}#2} }	
\newcommand{\proptime}{\tau}
\newcommand{\dist}[2]{ A^{#1}_{\ \, #2} }	
\newcommand{\F}[2]{ x^{#1}_{\,\, #2}\, }				
\newcommand{\devG}[1]{ \mathring{G}_{#1} }				
\newcommand{\devGmix}[2]{ \mathring{G}^{#1}_{\ #2} }				
\newcommand{\Cauchy}[1]{\sigma_{#1} }				
\newcommand{\Cauchymix}[2]{\sigma^{#1}_{\phantom{#1}#2} }		
\newcommand{\EMTD}[2]{\Sigma^{#1}_{\ \, #2} }				
\newcommand{\EMT}[2]{\mathcal{T}^{#1}_{\ \, #2} }				
\newcommand{\EMTGR}[2]{T^{#1}_{\ \, #2} }				
\newcommand{\gammix}[2]{\gamma^{#1}_{\ #2}} 	
\newcommand{\gam}[1]{\gamma_{#1}}		

\newcommand{\myxi}{\xi}
\newcommand{\g}{g}    

\newcommand{\LagrProjEff}{G}
\newcommand{\GeffContr}[1]{G^{#1}}
\newcommand{\Geffmix}[2]{G^{#1}_{\ #2}}
\newcommand{\Geff}[1]{G_{#1}}

\newcommand{\Kab}[1]{\kappa_{#1}}
\newcommand{\KAB}{\kappa}

\newcommand{\KMN}{\kappa}

\newcommand{\LED}{\mathcal{L}}	
\newcommand{\LEDGR}{\mathcal{L}^{\sGrm\sRrm}}	
\newcommand{\E}{\mathscr{E}}	

\newcommand{\scrU}{\mathscr{U}}	%
\newcommand{\thetas}{\theta_{\textrm sh}}
\newcommand{\thetah}{\theta_{\textrm h}}

\newcommand{\taus}{\tau_{\textrm sh}}
\newcommand{\tauh}{\tau_{\textrm h}}

\newcommand{\cs}{c_{\textrm sh}}
\newcommand{\ch}{c_{\textrm h}}

\newcommand{\alphah}{\alpha_{\textrm h}}
\newcommand{\uLag}{U}

\newcommand{\detA}{|A|}
\newcommand{\detxi}{|\myxi|}

\newcommand{\lapse}{\alpha}
\newcommand{\shift}{\beta}
\newcommand{\Lor}{\Gamma}

\newcommand{\entropy}{s}
\newcommand{\Entropy}{S}

\newcommand{\sA}{\mathsmaller A}
\newcommand{\sB}{\mathsmaller B}
\newcommand{\sC}{\mathsmaller C}

\newcommand{\sM}{\mathsmaller M}

\newcommand{\sV}{\mathsmaller V}

\newcommand{\sRrm}{{ \mathsmaller R}}
\newcommand{\sGrm}{{ \mathsmaller G}}

\begin{document}

\title{A new continuum model for general relativistic viscous heat-conducting 
media\footnote{Accepted for publication in \textit{Phil. Trans. R. Soc. A}}}

\author{
\normalsize
Evgeniy Romenski$^{1}$, Ilya Peshkov$^{2}$, Michael Dumbser$^{3}$ and Francesco 
Fambri $^{4}$}

\date{}
\maketitle


\let\thefootnote\relax\footnotetext{\textsuperscript{1} \textit{Sobolev 		
		Institute of Mathematics and Novosibirsk State University, Novosibirsk, 
		Russia
		\href{mailto:evrom@math.nsc.ru}{evrom@math.nsc.ru}}}
\let\thefootnote\relax\footnotetext{\textsuperscript{2} \textit{Laboratory of 
		Applied Mathematics, University of Trento, Trento, Italy, 
		\href{mailto:ilya.peshkov@unitn.it}{ilya.peshkov@unitn.it}}}
\let\thefootnote\relax\footnotetext{\textsuperscript{3} \textit{Laboratory of 
		Applied Mathematics, University of Trento, Trento, Italy, 
		\href{mailto:michael.dumbser@unitn.it}{michael.dumbser@unitn.it}}}
\let\thefootnote\relax\footnotetext{\textsuperscript{4} \textit{Max Planck 		
		Institute for Plasma Physics, Garching, Germany
		\href{mailto:francesco.fambri@ipp.mpg.de}{francesco.fambri@ipp.mpg.de}}}

\begin{abstract}
The lack of formulation of macroscopic equations for irreversible dynamics 
of viscous heat-conducting media compatible with the causality principle of 
Einstein's Special 
Relativity and the Euler-Lagrange structure of General Relativity is a 
long-lasting problem. In 
this paper, we propose a possible solution to this problem in the framework of SHTC equations. The 
approach does not rely on postulates of equilibrium irreversible thermodynamics but treats  
irreversible processes from the non-equilibrium point of view. Thus, each transfer process is 
characterized by a characteristic velocity of perturbation propagation in the 
non-equilibrium state, as well as by an intrinsic time/length scale of the dissipative dynamics.  
The resulting system of governing equations is 
formulated as a first-order system of hyperbolic equations with relaxation-type irreversible terms.
Via a formal asymptotic analysis, we demonstrate that classical transport 
coefficients such as 
viscosity, heat conductivity, etc. are recovered in leading terms of our theory as effective 
transport coefficients. 
Some numerical examples are presented in order to demonstrate the viability of the approach.
\end{abstract}


%


\section{Introduction}

The lack of formulations of macroscopic equations for irreversible dynamics 
of viscous heat-conducting and resistive media compatible with the causality 
principle of Einstein's special relativity and the Euler-Lagrange structure of 
general relativity (GR) is a long lasting problem 
\cite{Hiscock1983,Hiscock1985,RezzollaZanottiBook}. 
In 
this paper, we propose a possible solution to this problem in the framework of 
Symmetric Hyperbolic and Thermodynamically Compatible (SHTC) equations 
\cite{SHTC-GENERIC-CMAT,God1961,GodRom1995,Godunov1996,Rom1998,GodRom2003}. 
Such 
an approach is not relying on postulates of equilibrium irreversible 
thermodynamics but treats  
irreversible processes from the non-equilibrium standpoint. Thus, each transfer 
process is 
characterized by a characteristic velocity of perturbation propagation $ c_{\textrm ch} $ in the 
non-equilibrium state as well as by an intrinsic time scale $ \tau $ of the 
dissipative (irreversible) dynamics.  
The resulting system of governing equations is 
formulated as a first-order system of hyperbolic equations with 
relaxation-type irreversible terms and thus causal by construction. 
Via a formal asymptotic analysis, we demonstrate that classical transport 
coefficients such as 
viscosity and heat conductivity are recovered in leading terms of our theory 
as effective 
transport coefficients of the form $ \sim \tau c_{\textrm ch}^2 $. 
Some numerical examples will be presented in order to demonstrate the 
viability of the approach.

The overall time evolution described by SHTC equations is split into two parts, reversible 
and irreversible. The reversible part (all the differential terms) comprises most of the 
mathematical structure of the governing equations. This part is associated with the most 
non-equilibrium state of the system, i.e. for the relaxation parameter approaching infinity, $ \tau 
\to \infty $. On the other hand, when $ \tau \to 0$, the system is driven 
towards 
the 
global thermodynamic equilibrium which is described by a system with a reduced 
structure (the Euler equations of an ideal inviscid fluid).

The reversible part conserves both the energy and the entropy and admits a variational formulation 
and thus, it is compatible with the Euler-Lagrange structure of the Einstein field equations of 
GR. An important feature of our approach is that the variational principle is 
formulated in the Lagrangian reference frame, while the final form of the governing equations is 
obtained after the passage from the Lagrangian to the Eulerian frame. This is discussed in 
Sec.\,\ref{sec.reversible}. Finally, we recover a covariant form of the 
Eulerian equations, and the 
corresponding $3+1$ split ('\emph{Valencia-type}') formulation \cite{Anton06,Anton2010,Marti2015}, which is also well suited for carrying out numerical experiments for a first and direct validation of the proposed theory. 

The irreversible part, which rises the entropy (second law of thermodynamics), 
is 
represented by algebraic relaxation-type source terms $ \sim \tau^{-1} $, and 
can be 
viewed as gradients (with respect to the state variables) of a 
dissipation potential \cite{SHTC-GENERIC-CMAT}, see Sec.\,\ref{sec.full.model}. 

The relaxation character of the irreversible part, in particular, imposes a challenge for the 
numerical methods in the 
near equilibrium (diffusive) regime $ \tau \ll 1 $. Therefore, in the numerical simulations, we 
rely on the so-called ADER approach \cite{toro4,DumbserEnauxToro,ADERGRMHD} which can be used to obtain 
the numerical solution in all regimes: near equilibrium $ \tau \ll 1 $, non-equilibrium 
$ \tau \gg 1 $, and intermediate $ \tau \sim 1 $, see \cite{DPRZ2016,DPRZ2017}. 

The SHTC approach was not developed as a non-equilibrium thermodynamics theory, but grew up from 
studying the admissible mathematical structure of macroscopic equations initiated by Godunov in 
\cite{God1961}. Such a structure should simultaneously
guarantee consistency with the principles of thermodynamics and the causality principle, 
and also should have good mathematical properties such as well-posedness of the initial value problem (Cauchy problem),  
which is in particular obligatory for a system of evolutionary equations to be solved numerically. 
Nevertheless, the SHTC approach shares many common features with other non-equilibrium 
thermodynamics theories. For example, it has been shown recently \cite{SHTC-GENERIC-CMAT} that the 
SHTC equations admit a Hamiltonian formulation via Poisson brackets and thus, it can be seen as a 
particular realization of the GENERIC formulation of non-equilibrium 
thermodynamics \cite{Ottinger-book,PKG-Book2018}. Moreover, as many 
theories, e.g. GENERIC, Extended Irreversible Thermodynamics (EIT) \cite{EIT2010}, Rational 
Extended Thermodynamics (RET) \cite{MullerRuggeri1998}, the SHTC approach intends to identify new 
macroscopic fields (state parameters) and find evolution equations governing them in order to 
approach more and more non-equilibrium regime. The physical meaning of the new 
state parameters 
might be very different though from those used in RET and EIT, e.g. flux-type 
quantities in EIT and RET and density-type 
quantities in SHTC and GENERIC. Additionally, similar to RET, special care is given to the 
hyperbolicity  of the governing equations.

We note that a distinguishing feature of the SHTC and Hamiltonian GENERIC 
approach is the 
exceptional role of the energy potential in formulating the governing equations. Indeed, the 
variational nature of the SHTC equations and the Hamiltonian nature of GENERIC imply a unique role of 
the Lagrangian and the Hamiltonian, respectively, which act as generating potentials (they generate the entire structure of the reversible part of the governing PDE system) and are intimately connected 
with the energy of the system. On the other hand, in theories such as EIT and RET, the role of the 
generating potential is given to the entropy. In SHTC and GENERIC, the entropy potential, which is 
intimately connected to the \textit{dissipation potential} 
\cite{SHTC-GENERIC-CMAT}, is only responsible for generating the 
irreversible part of the time evolution.

Concerning the existing approaches to relativistic fluid mechanics, it is worthwhile 
to mention that 
there are also many 
theories for causal 
relativistic dissipation which can be roughly divided into two classes. The first 
class consists of models which are presented by mixed-order (first- and second-order) PDE systems 
such as 
the original covariant reformulation of the Navier-Stokes-Fourier equations 
	put forward by 
	Eckart \cite{Eckart1940} and Landau and 
	Lifshitz~\cite{LandauLifshitzHydro} and which have been shown later to be \textit{acausal} 
	and unstable \cite{Hiscock1985}. Recently, the causality and stability issues of the 
	Eckart-Landau theory was addressed by several authors  
	\cite{Van2012,Freistuhler2017,Freistuhler2018,Bemfica2018,Bemfica2019a}. Thus, in contrast to 
	their 
	predecessors \cite{Eckart1940,LandauLifshitzHydro}, V\'an and Bir\'o used the Lagrange 
	multiplier approach to obtain a stable formulation in \cite{Van2012}, Freist\"uhler and
	Temple \cite{Freistuhler2017,Freistuhler2018} derived a causal and stable formulation 
	which has a locally well-posed initial value problem and has a close connection with the 
	second-order 
	symmetric hyperbolic systems in the sense of 
	Hughes-Kato-Marsden \cite{Hughes1977} and which is referred to as mixed-order symmetric 
	hyperbolic system\footnote{In this work, we consider only first-order systems 
	and the symemtric hyperbolicity is understood with respect to such systems, i.e. in the sense 
	of Friedrichs 
	\cite{Friedrichs1958,God1961}.} by the authors. Recently, Bemfica and co-authors 
	\cite{Bemfica2018,Bemfica2019a} proposed 
	another causal, stable and locally well-posed extension of the Eckart-Landau theory, first in 
	the case 
	of conformal fluids \cite{Bemfica2018} and later in the general case \cite{Bemfica2019a} which 
	also can be derived from the relativistic Boltzmann kinetic equation via a perturbative
	expansion technique developed in \cite{Denicol2011}.
	See a more thorough review of the aforementioned approaches in \cite{Bemfica2018}.
The second class of models consists of those which are presented by first-order hyperbolic PDE 
systems 
such as 
M\"uller-Israel-Stewart theory 
\cite{Mueller1966PhD,Muller1967,Israel1976,Stewart1977} which was originally established as a 
phenomenological extended thermodynamic theory but later it was shown \cite{MullerRuggeri1986} 
that 
it can benefit from the
connection with the relativistic Boltzmann equation via the moment expansion method 
of Grad \cite{Grad1949}. There are various
state of the art extensions and modifications of the original M\"uller-Israel-Stewart theory, e.g. 
see
\cite{RezzollaZanottiBook,Romatschke2010,MullerRuggeri1986,
	HuovinenMolnar2009,Denicol2013,Molnar2012,Pennisi2017}. Also,  a quite general class of 
	divergence-type theories \cite{Reula2018,RezzollaZanottiBook} can be related to the first-order 
	PDE models. In particular, the divergence-type models are known to be symmetric hyperbolic by 
	contraction due to the Godunov-Boillat theorem \cite{SHTC-GENERIC-CMAT}.
As it is 
shown on the example 
of viscous dissipation in \paperI, the theory proposed in this paper is 
equivalent to the 
M\"uller-Israel-Stewart theory up to the first-order terms in 
the out of equilibrium formal Chapman-Enskog
expansion in the small relaxation parameter. However, the higher order terms 
are different as 
they  differ within various versions of the M\"uller-Israel-Stewart theory 
\cite{Molnar2012} 
depending on 
the selected closure relations.

Let us finally name the two most important features that should 
help the reader to distinguish our approach within the aforementioned
approaches to relativistic dissipative fluid mechanics. First of all, an attractive feature 
of our theory
is that it admits a variational formulation so that the matter 
energy-momentum is equivalent to 
the GR canonical matter energy-momentum tensor. In other words, coupling of the matter and gravity 
occurs in a natural way, that is in contrast to the existing approaches, the 
energy-momentum tensor of our 
theory does not have to be added in an \textit{ad hoc} manner to the Einstein 
field equations but emerges there as the Noether current via the variation of the Hilbert-Einstein 
action. The same can be said for coupling of matter and electromagnetic force as in the nonlinear 
electrodynamics of moving medium 
\cite{DPRZ2017}. Additionally, due to the variational nature of the SHTC equations discussed here, 
the governing equations have a certain structure conditioned by the canonical structure of the 
Euler-Lagrange equations. This is convenient for designing of numerical methods and in particular, 
for developing of structure preserving methods. Let us also remark that despite the paramount role 
of the variational principle in relativistic 
physics, to the best of our knowledge, there were no 
much 
attempts to employ variational principle for deriving equations for relativistic dissipative 
continuum mechanics apart from the works \cite{Carter1991,Andersson2015,Andersson2007a}. However, 
at the end of the day, the 
viscosity law is postulated but not derived in these papers.

The second distinguishing feature of our approach is that it provides a \emph{unified} 
description of fluids and solids \cite{HPR2016,DPRZ2016,HYP2016,Busto2019}. In contrast to the 
existing first-order hyperbolic approaches which rely on the kinetic theory of gases or parabolic 
theories which rely on the phenomenological Navier-Stokes-Fourier constitutive laws, our theory is 
not restricted by 
applications to only fluids but can be applied simultaneously to both relativistic liquids and 
solids (e.g. the star 
interior, the outer crust of the neutron stars~\cite{Chamel2008}).

\section{Formulation of reversible equations}\label{sec.reversible}

In this section, we give a variational formulation of the reversible part of the 
time evolution for relativistic viscous and elastoplastic heat-conducting media. 
As it will be clear from what follows, our formulation involves 
two manifolds, the 
matter manifold and spacetime manifold. It  is therefore not obvious on which manifold of the two 
one has to formulate the variational principle. In fact, the Euler-Lagrange equations are the same 
regardless whether the variations are performed on matter or spacetime manifold. However, the 
Euler-Lagrange equations are not the only equations we need in our theory. We also need the 
so-called 
\emph{integrability conditions} \eqref{eqn.motion.Eul}$ _2 $ and \eqref{eqn.GRHC}$ _2 $ which form 
the 
evolution 
equations for the principal fields of the 
theory, which are the distortion field and the thermal impulse. It appears, that these equations 
cannot be rigorously derived if one works on the spacetime manifold (Eulerian frame) but they can 
only be obtained in an \textit{ad hoc} manner. On the other hand, the formulation of the 
variational principle on the matter manifold (Lagrangian frame) allows us to obtain all 
equations in a rigorous mathematical way. Therefore, in what follows, we formulate the variational 
principle in the matter manifold (Lagrangian frame) and then transform the Euler-Lagrange equations 
and integrability 
conditions to the Eulerian frame associated with the spacetime manifold. In particular, we show 
that the matter energy-momentum tensor obtained in such a way is equivalent to the GR matter 
energy-momentum obtained in the standard way, i.e. by varying the action with respect to the 
spacetime metric $ g_{\mu\nu} 
$. 

Also, note that the 
governing equations are formulated in such a way that the 
Lagrangian density is left unspecified and has to be provided by the user in 
order to close the equations. Such a closure may depend on a particular 
application. This emphasizes an exceptional role of the energy potential in the 
formulation of the reversible equations, similar to the GENERIC formulation 
\cite{Ottinger-book,PKG-Book2018}.

\subsection{Lagrangian equations of motion}\label{sec.Var.Lagr}
In continuum mechanics, the motion of a continuous medium can be viewed as an embedding of the $ 4d $
matter-time manifold (4-continuum) $ \mattertime $ with a Lorentzian metric $ \Kab{ab} $ in the 
$4d$ spacetime 
$\spacetime$ with a general Lorentzian metric $ g_{\mu\nu} $, see \paperI.
This embedding
can be 
described in \emph{Lagrangian 
coordinates}  $ \myxi^a $ associated with $ \mattertime $, i.e. the coordinates that are 
\emph{comoving} and \emph{co-deforming} with 
the medium, or, alternatively, in generic non-comoving coordinates $ x^\mu $ associated with $ 
\spacetime $, that are usually 
called \emph{Eulerian} coordinates for convenience. Sometimes the Lagrangian 
coordinates are also named \emph{material} coordinates, but in this work we 
prefer the first against the latter.
The embedding implies that the following \textit{one-to-one} relation
\begin{equation}\label{eqn.motion.laws1}
	x^\mu = x^\mu(\myxi), \qquad \myxi^a = \myxi^a(x), \quad \mu,a=0,\ldots,3 
\end{equation}
holds between Lagrangian and Eulerian coordinates, also called \textit{motion}. 
In this work, by convention, we refer general Greek indexes to the non-comoving 
(Eulerian) system of coordinates, while lowercase Latin indexes $a,b,c =0,1,2,3$ to the 
comoving 
(Lagrangian) coordinates. Additionally, capital Latin letters $ \sA, $ $ \sB $, 
$ \sC=1,2,3 $ 
refer to the three purely spatial Lagrangian material coordinates, e.g. $ \myxi^\sA $.

We further specify the Lagrangian coordinates $ \myxi^a $ by assuming that the 
three scalars 
$\myxi^{\sA} $ \textit{label} 
the 
matter particles and hence label the particle worldlines, while $ \myxi^0 
:=\proptime  $ is defined 
to 
be the { matter} \textit{proper time}, that is the time of 
the Lagrangian observer as measured from his comoving clock, i.e.
\begin{equation}\label{eqn.proper.time}
-\rmd \proptime ^2 = g_{\mu\nu} \rmd x^\mu \rmd x^\nu.
\end{equation}

In the Lagrangian formalism, the motion \eqref{eqn.motion.laws1}, as seen by an 
Eulerian observer, represents deformation of the medium and hence, the gradients 
of the motion, also called \textit{configuration gradients} in the relativistic 
elasticity literature, e.g. \cite{Kijowski1992,Kijowski1998,Wernig-Pichler2006,
	Broda2008,Gundlach2012},
\begin{equation}\label{eqn.Jacobians}
	\F{\mu}{a}(\myxi) := \frac{\pd x^\mu}{\pd \myxi^a}, \qquad 
	\cgrad{a}{\mu}(x) := 
	\frac{\pd \myxi^a}{\pd x^\mu} 
\end{equation}
play a central role \paperI. In particular, the 4-velocity of the material elements with 
respect to the Eulerian coordinate system $ x^\mu $ is defined as the first column of the 
configuration gradient $ \F{\mu}{a} $
\begin{equation}\label{eqn.4vel.def}
u^\mu := \F{\mu}{0} = \frac{\pd x^\mu}{\pd \myxi^0} = \frac{\pd x^\mu}{\pd \proptime },
\end{equation}

Thus, in the absence of other material fields 
(which will be introduced later), it is implied that the material Lagrangian 
density 
$ \tilde{\Lag} $ explicitly depends on the configuration gradient $ \F{\mu}{a} $
\begin{equation}\label{eqn.Lagrangian}
\tilde{\Lag}\left(\myxi^a, x^\mu(\myxi),\F{\mu}{b}(\myxi)\right) =
\Lag\left(\F{\mu}{b}(\myxi)\right)
\end{equation}
and does not explicitly depend on the unknowns $ \myxi^a $ and the potentials 
$ x^\mu(\myxi) $. Hence, the first variation of 
the action $\Act = \int \Lag\, \rmd 
\boldsymbol{\myxi}$ with respect to $ \delta x^\mu $ gives the Euler-Lagrange 
equation
\begin{subequations}\label{eqn.motion.Lag}
	\begin{equation}\label{eqn.Euler-Lagrange}
		\pd_a\,\Lag_{\F{\mu}{a}} = 0,
	\end{equation}
	where  we introduced the notation $ 
	\Lag_{\F{\mu}{a}} = \frac{\pd \Lag}{\pd \F{\mu}{a}} $, and  the Einstein 
	convention of summation over repeated indexes is assumed.
	There are only 4 conservation laws in \eqref{eqn.Euler-Lagrange} for 16 
	unknowns $ \F{\mu}{a} $. The remaining 12 equations are indeed hidden within
	 the \textit{integrability conditions} of any configuration gradient, i.e.
	\begin{equation}\label{eqn.integrability}
		\pd_b\, \F{\mu}{a} - \pd_a\, \F{\mu}{b} = 0, \quad a\neq b.
	\end{equation}
\end{subequations}
In fact, system \eqref{eqn.integrability} consists of $24$  equations, where 12 of them 
are effective \textit{evolution} equations, i.e. those for $ a,b = 0 $, while the other 12 are the
so-called \textit{involution constraints}, e.g. see~\cite{GRGPR,SHTC-GENERIC-CMAT}, that 
are pure spatial constraints, conserved along the particle trajectories.

\subsection{Eulerian equations of motion}


Existence of the Lagrangian coordinates $ \myxi^a $ is a mathematical 
idealization and their practical use for general fluid-like motion is usually very 
problematic. Therefore, our unified approach to fluids and solids \cite{HPR2016,DPRZ2016} relies on 
the 
reformulation of governing equations \eqref{eqn.motion.Lag} in the Eulerian 
frame and later, on the replacement of 
the integrable (global) deformation field $ \cgrad{a}{\mu} $ by the 
non-integrable (local) 
distortion field $ \dist{a}{\mu} $ which can be seen as a local basis tetrad 
(or non-holonomic basis tetrad). Therefore, in this section, we formulate 
equations 
of motion in the Eulerian frame which are obtained from the Lagrangian 
governing equations \eqref{eqn.motion.Lag} by changing the unknowns $ \myxi^a 
\to x^\mu $, see details in \paperI.

Thus, after a sequence of equation transformations, equations 
\eqref{eqn.motion.Lag} read
\begin{equation}\label{eqn.motion.Eul}
-\nabla_\nu(\detxi \, \cgrad{b}{\mu} \Lambda_{\cgrad{b}{\nu}}) = 0, 
\qquad
u^\nu (\nabla_\nu\cgrad{a}{\mu} - \nabla_\mu\cgrad{a}{\nu}) = 0,
\end{equation} 
where $ \nabla_\nu $ denotes a covariant derivative associated with the 
symmetric Levi-Civita connection of GR, and the first equation represents the 
conservation of the energy-momentum tensor-density $ \EMTD{\nu}{\mu} $ of our 
theory
\begin{equation}
\EMTD{\nu}{\mu} :=-\detxi\, \cgrad{\mu}{a}\Lambda_{\cgrad{\nu}{a}} = \detxi\, 
\F{\nu}{a}\Lambda_{\F{\mu}{a}}, \qquad \detxi = \det(\cgrad{a}{\mu}).
\end{equation}
After introducing the Eulerian counterpart $ \LED $ of the material Lagrangian $ 
\Lag $:
\begin{equation}
\LED := \detxi \Lambda,
\end{equation}
the energy-momentum reads
\begin{equation}\label{eqn.divEnMom1}
-\EMTD{\nu}{\mu} = \cgrad{a}{\mu} \LED_{\cgrad{a}{\nu}} - \LED \delta^{\nu}_{\ \mu}, \qquad 
\EMT{\nu}{\mu} := \EMTD{\nu}{\mu}/\sqrt{-g}, 
\qquad g = \det(g_{\mu\nu}).
\end{equation}
In particular, it is shown in \paperI\ that in the absence of other material 
and 
electromagnetic fields, $ \EMT{\nu}{\mu} $ can be written in a conventional 
form 
\begin{equation}\label{energy-momentum.conv}
-\EMT{\nu}{\mu} = \E u^\nu u_\mu + p h^\nu_{\ \mu} + \sigma^\nu_{\ 
	\mu}, \qquad h^\nu_{\ \mu} := \delta^\nu_{\ \mu} + u^\nu u_\mu,
\end{equation}
where $ \E = \LED/\sqrt{-g} $ is the total energy, $ p := \rho \E_\rho -\E $ 
is 
the isotropic pressure, $ 
\rho $ is the rest mass density, $ \sigma^\nu_{\ \mu} = 
\cgrad{a}{\mu}\E_{\cgrad{a}{\nu}} $ is the anisotropic part of the 
energy-momentum, $ h^\nu_{\ \mu} $ is the spatial projector.

The complete system of Eulerian governing equations for determining the 
16 unknown fields $ \cgrad{a}{\mu} $ reads
\begin{equation}\label{eqn.motion.Eul1}
-\nabla_\nu(\cgrad{a}{\mu} \E_{\cgrad{a}{\nu}} - \E \delta^{\nu}_{\ \mu}) = 
0, 
\qquad
u^\nu (\nabla_\nu\cgrad{a}{\mu} - \nabla_\mu\cgrad{a}{\nu}) = 0.
\end{equation}

Finally, we note that, as discussed in detail in \cite{GRGPR,PRD-Torsion2019}, 
the developed 
approach is geometrical in nature, in which the matter is considered as a 
non-Riemannian manifold in which the geometry is determined by the 
distortion field $ \dist{a}{\mu} $ (non-holonomic frame field), which replaces 
the holonomic tetrad $ \cgrad{a}{\mu} $ and plays the role of moving Cartan 
frames. In such a geometrical framework, the 4-velocity $ u^\mu $ and the rest 
mass current $ j^{\mu}:=\sqrt{-g}\rho u^\mu $ are collinear by construction 
\cite{GRGPR}. In other words, the proposed  approach can be viewed as a 
generalization of Eckart's choice of the 4-velocity (Eckart's frame), see 
\cite{Eckart1940,Van2008,RezzollaZanottiBook}.

\paragraph{Consistency with GR.}
As it was shown in \paperI, our energy-momentum tensor \eqref{eqn.divEnMom1} 
agrees well with the 
canonical matter energy-momentum tensor-density of GR (the source term in the 
Einstein field equations) which reads as
\begin{equation}\label{energy.momentum.GR}
\sqrt{-g}\,\EMTGR{\nu}{\mu} := -g^{\nu\lambda} \frac{\pd \LEDGR}{\pd g^{\lambda\mu}} = 
g_{\mu\lambda} 
\frac{\pd 
\LEDGR}{\pd g_{\lambda\nu}} = \sqrt{-g}\left(2 g_{\mu\lambda}\E_{g_{\lambda\nu}} - \E 
\delta^{\nu}_{\ \mu} \right),
\qquad
\LEDGR := \sqrt{-g} \E,
\end{equation}
where $ \E $ is the total energy.
In particular, if we assume $ \LED = \LEDGR $, it was shown in \paperI\ that 
\begin{equation}\label{GR.SHTC.energymom}
-\EMT{\nu}{\mu} = 
\cgrad{a}{\mu} \E_{\cgrad{a}{\nu}} - \E \delta^{\nu}_{\ \mu} = 
-\left(2 g_{\mu\lambda}\E_{g_{\lambda\nu}} - \E \delta^{\nu}_{\ \mu} \right) = 
-\EMTGR{\nu}{\mu}.
\end{equation}
This is an important result, saying that our way of formulating the 
variational principle, which consists in defining the action in the 
matter-time manifold $ 
\mattertime $ and varying the 
action with 
respect to the configuration gradient $ \cgrad{a}{\mu} 
$, gives the energy-momentum tensor \eqref{GR.SHTC.energymom} equivalent to 
the conventional GR 
matter energy-momentum, which is obtained if the action is defined in the 
spacetime $ \spacetime $ and the variation is performed with 
respect to the spacetime metric $ g_{\mu\nu} $. 

Equality \eqref{GR.SHTC.energymom} also guarantees that  
the GR energy-momentum tensor $ \EMTGR{\nu}{\mu} $ coincides identically with the SHTC 
energy-momentum $ \EMT{\nu}{\mu} $ for arbitrary energy potential $ \E $ even 
in the presence of extra fields, e.g. the thermal impulse needed
for the heat conduction formulation.

In the end, we note that the proposed continuum theory does not 
rely on any specific 
assumptions 
of GR, e.g. \textit{Einstein's equivalence principle}, and in fact, it can be 
viewed as a theory 
whose main feature is the causal description of dissipative phenomena, see 
Sec.\ref{sec.therm.consistency}, i.e. it is compatible with the  special 
relativity theory. 
Nevertheless, a side effect of using  the configuration gradient $ 
\cgrad{a}{\mu} $ (the frame 
field) as the principal field is that the SHTC energy-momentum tensor $ \EMT{\nu}{\mu} $ is 
equivalent to 
the GR canonical matter energy-momentum tensor $ \EMTGR{\nu}{\mu} $ for arbitrary background 
Lorentzian metric $ 
g_{\mu\nu} $. 
Moreover, due to their specific structure, the governing equations are also 
generally 
covariant, e.g. see \paperI\ and eqs.\eqref{euler.eq.J4} and 
\eqref{euler.eq.J5} below, for the
torsion-less 
spacetimes.

\subsection{Heat conduction} \label{sec.Heat}

In this Section, we give a variational formulation for the reversible part of 
the relativistic version of 
the SHTC heat conduction equations \cite{Rom1998,SHTC-GENERIC-CMAT}.

As it was 
shown in \cite{SHTC-GENERIC-CMAT}, \textcolor{ForestGreen}{non-relativistic} 
SHTC equations can be 
viewed as a 
particular realization of the GENERIC formulation for non-equilibrium 
thermodynamics. It is therefore not surprising that the same heat conduction 
equations were also proposed by \"Ottinger 
\cite{Ottinger1998,Ottinger1998b,Ottinger2019} within the GENERIC approach 
\textcolor{ForestGreen}{in the relativistic 
context}.

\subsubsection{Lagrangian equations.}\label{Lagrangian}

In the Lagrangian frame, let us consider a scalar potential $ \pot(\myxi^a) $ 
and the action integral 
\begin{equation}\label{heat.action}
	\Act = \int\Lambda(\pot,\pd_a J)\rmd \bm{\myxi}.
\end{equation}
We assume that the Lagrangian $ \Lambda $ does not depend explicitly on the potential $ \pot $ itself,  
but only on its first derivatives $ \pot_a :=\pd_a \pot $. In analogy with non-relativistic 
equations \cite{DPRZ2016,DPRZ2017,SHTC-GENERIC-CMAT}, vector $ \pot_a $ is 
called relativistic
\textit{thermal impulse}. Furthermore, we shall single out the 
zeroth component $ \pot_0 := -T $ of the gradient
 $ \pot_a $,
where, as will become clear later, $ \Temp $ 
can be identified as the temperature in the medium.

The first variation of the action \eqref{heat.action} with respect to $ 
J $ gives the 
Euler-Lagrange 
equation
\begin{equation}\label{heat.EL}
	\pd_a \Lambda_{\pot_a} = 0, \qquad \partial_b \pot_a - \partial_a \pot_b = 0,
\end{equation}
which are accompanied by the \textit{integrability condition} (second 
equation). The first equation in \eqref{heat.EL} will be associated with the 
entropy evolution equation and its more conventional form will be unveiled 
later. The second equation\footnote{Note that within 
\eqref{heat.EL}, there are only 4 evolution equations for 4 unknowns $ \pot_a $ 
and 3 pure spatial constraints $ \pd_\sB \pot_\sA - \pd_\sA \pot_\sB = 0$.}
will be used to evolve the three spatial components of the thermal impulse.

\subsubsection{Eulerian equations.}

We now rewrite the Lagrangian equations~\eqref{heat.EL} in the Eulerian frame $ x^\mu $.
Thermal impulse $ \pot_a $ and Lagrangian $ \Lambda $ transform as
\begin{equation}\label{euler.eq.change.J}
\pot_\mu = \cgrad{a}{\mu} \pot_a, \qquad \LED(\pot_\mu) = \detxi 
\Lambda(\pot_a).
\end{equation}
In particular, keeping in mind that $ u^\mu = \F{\mu}{0} $, we have 
\begin{equation}\label{euler.eq.temp}
u^\mu \pot_\mu = -T.
\end{equation}

Furthermore, it follows from \eqref{euler.eq.change.J} that $ \detxi 
\F{\mu}{a}\Lambda_{\pot_a} =  \LED_{\pot_\mu} $ and hence, 
 using 
the identity $ \pd_\mu (\detxi \F{\mu}{a}) = 0 $ and that $ \pd_a = 
\F{\mu}{a}\pd_a $, one can rewrite 
\eqref{heat.EL}$ _1 $ as
\begin{equation}
  0 = \detxi\,\F{\mu}{a}\pd_\mu  \Lambda_{\pot_a} =  
	\pd_\mu (\detxi\,\F{\mu}{a}\Lambda_{\pot_a})  =
	  \pd_\mu \LED_{\pot_\mu},
\end{equation}
that is the vector-density $ \LED_{\pot_\mu} $ conserves both ordinarily and 
covariantly:
\begin{equation}\label{euler.eq.entropy}
	\pd_\mu \LED_{\pot_\mu} = 0, \qquad \text{and}\qquad \nabla_\mu\LED_{\pot_\mu} = 0.
\end{equation}

Let us now transform equation \eqref{heat.EL}$ _2 $. One may write (we consider only three 
genuine evolution equations, but not the spatial constraints $ \pd_\sB \pot_\sA - \pd_\sA \pot_\sB = 0 $)
\begin{equation}\label{euler.eq.J1}
	\pd_\tau \pot_b - \pd_b \pot_0 = \pd_\tau (\F{\mu}{b}\pot_\mu) - \pd_b (u^\mu 
	J_\mu) = u^\lambda\pd_\lambda (\F{\mu}{b}\pot_\mu) - \F{\lambda}{b}\pd_\lambda 
	(u^\mu J_\mu).
\end{equation}

Then, using \eqref{euler.eq.temp} and that $ \pd_\lambda 
\F{\mu}{b}=-\F{\eta}{b}\F{\mu}{a}\pd_\lambda 
\cgrad{a}{\eta} $, we can write \eqref{euler.eq.J1} as
\begin{equation}\label{euler.eq.J3}
	u^\lambda \pd_\lambda \pot_\nu - u^\lambda \pot_\mu 
	\F{\mu}{a}\pd_\lambda\cgrad{a}{\nu} +
	\pd_\nu\Temp = 0,
\end{equation}
which using the evolution equation for the configuration gradient $ 
u^\lambda\pd_\lambda\cgrad{a}{\nu} + \cgrad{a}{\lambda}\pd_\nu u^\lambda = 0$ (see \paperI), can be 
rewritten as
\begin{equation}\label{euler.eq.J4}
	u^\lambda \pd_\lambda \pot_\nu + \pot_\mu\pd_\nu u^\mu = -\pd_\nu \Temp.
\end{equation}
Because the left hand-side is the Lie derivative of $ \pot_\mu $ along the 4-velocity, while the 
right hand-side is the derivative of the proper scalar $ \Temp = -u^\mu 
\pot_\mu $, we may replace the ordinary 
derivatives with the covariant ones:
\begin{equation}\label{euler.eq.J5}
	u^\lambda \nabla_\lambda \pot_\nu + \pot_\mu\nabla_\nu u^\mu = 
	-\nabla_\nu\Temp.
\end{equation}
Using \eqref{euler.eq.temp}, this equation can be equivalently written as $ 
u^\lambda(\nabla_\lambda \pot_\nu - \nabla_\nu \pot_\lambda) = 0 $. Thus, the 
Eulerian form of the relativistic heat conduction equations (in the absence of the irreversible 
processes) reads 
\begin{equation}\label{eqn.GRHC}
	\nabla_\mu\LED_{\pot_\mu} = 0,
	\qquad
	u^\lambda(\nabla_\lambda \pot_\nu - \nabla_\nu \pot_\lambda) = 0.
\end{equation}

\subsubsection{Energy-Momentum and Thermal Stress.}

We now demonstrate that the energy-momentum tensor has a contribution due to 
thermal impulse $ \pot_\mu $. Thus, during the Lagrange-to-Euler transformation the energy-momentum 
tensor density $ \EMTD{\nu}{\mu} = \sqrt{-g} \EMT{\nu}{\mu}$ 
\eqref{eqn.divEnMom1} transforms as
\begin{equation*}
\EMTD{\nu}{\mu} = \detxi\, \F{\nu}{a} 
\left (\Lambda(\F{\lambda}{b},\pot_b)\right )_{\F{\mu}{a}} 
= 
- \detxi\, \cgrad{a}{\mu} \left ( \detxi^{-1} 
\LED(\cgrad{b}{\lambda},\cgrad{b}{\lambda}\pot_b)\right )_{\cgrad{a}{\nu}} 
= 
\LED \delta^{\nu}_{\ \mu} - \cgrad{a}{\mu} \LED_{\cgrad{a}{\nu}} - 
\pot_\mu \LED_{\pot_\nu}.
\end{equation*}
Therefore, in the presence of heat conduction, the energy-momentum tensor $ 
\EMT{\nu}{\mu} = \EMTD{\nu}{\mu}/\sqrt{-g} $ is
\begin{equation}\label{heat.EM}
	-\EMT{\nu}{\mu} = \cgrad{a}{\mu} \E_{\cgrad{a}{\nu}} + \pot_\mu 
	\E_{\pot_\nu} - \E \delta^\nu_{\ \mu},
\end{equation}
where the term $ \pot_\mu \E_{\pot_\nu} $ can be called the \textit{thermal 
stress}.

\subsubsection{Entropy.}

In this section, we introduce an entropy of our theory and rewrite the conservation 
law \eqref{eqn.GRHC}$ _1 $, which is in fact the entropy conservation law, in a 
conventional form.
For brevity, we shall not consider the dependence of the 
energy potential on $ \cgrad{a}{\mu} $ in this section.

First, we introduce the spatial projection $ \potw_\mu $ of the thermal impulse 
$ 
\pot_\mu $
\begin{equation}\label{proj.J}
	\potw_\mu := h^\lambda_{\ \mu}\pot_\lambda = \pot_\mu - \Temp u_\mu.
\end{equation}

Let us also define total energy potentials (at the moment we use symbol $ E $, 
while symbol $ \E $ will appear later on, after a Legendre transform in 
\eqref{entropy.temperature})
\begin{equation}\label{energies}
	E(\pot_\mu)=\ETJ(\Temp,J) = \ETW(\Temp,\potw),
\end{equation}
where $ E = \LED/\sqrt{-g} $, $	\Temp = -u^\mu\pot_\mu$, $\pot := \pot_\mu 
\pot^\mu$, $ \potw := \potw_\mu \potw^\mu = J + \Temp^2$,
and, in particular,
\begin{equation}
E_{\pot_\mu} = -\ETW_\Temp u^\nu + \ETW_{\potw_\lambda}\frac{\pd 
\potw_\lambda}{\pd 
	\pot_\nu} = 
-\ETW_\Temp u^\nu + \ETW_{\potw_\lambda}h^\nu_{\ \lambda} =
-\ETW_\Temp u^\nu + \ETW_{\potw_\nu}.
\end{equation}


Eventually, the entropy $ \entropy $ can be introduced conventionally, i.e. as 
the 
Legendre 
conjugate to the temperature $ \Temp $. Thus, we introduce a new potential $ 
\E(\entropy,\potw_\mu) 
= \Temp \ETW_\Temp - \ETW $ and hence, one has
\begin{equation}\label{entropy.temperature}
	\entropy := \ETW_\Temp,
	\qquad
	\Temp = \E_\entropy,
	\qquad
	\E_{\potw_\mu} = -\ETW_{\potw_\mu}.
\end{equation}
After that, conservation law~\eqref{eqn.GRHC}$ _1 $ reads
\begin{equation}
\nabla_\mu E_{\pot_\mu} = \nabla_\mu(-\ETW_\Temp u^\mu + \ETW_{\potw_\mu}) =
-\nabla_\mu( \entropy u^\mu + \E_{\potw_\mu}),
\end{equation}
which represents the reversible part of the entropy evolution. Note that the 
non-advective part $ \E_{\potw_\mu} $ of the entropy current is orthogonal to 
the 
4-velocity 
\begin{equation}\label{orthogon}
\E_{\potw_\mu} u_\mu = 0
\end{equation}
because $ \E_{\potw_\mu} = -\ETW_{\potw_\mu} = -2 \ETW_\potw \potw^\mu$ and $ 
\potw^\mu u_\mu = 0 $.
In terms of the thermal impulse $ \potw_\mu $, eq.\,\eqref{eqn.GRHC}$ _2 $ reads
\begin{equation}\label{w.PDE}
u^\lambda(\nabla_\lambda \potw_\mu - \nabla_\mu \potw_\lambda) + h^\lambda_{\ 
\mu} Q_\lambda = 0,
\end{equation}
where $ Q_\mu := \nabla_\mu T + T u^\lambda \nabla_\lambda u_\mu $ is the 
auxiliary temperature gradient vector, e.g. see 
\cite{Ottinger1998,RezzollaZanottiBook}. However, in the numerical simulation, 
it is more convenient to use equation \eqref{eqn.GRHC}$ _2 $ for $ \pot_\mu $ 
and then compute $ \potw_\mu $ using \eqref{proj.J} 
because, after the $ 3+1 $ split, \eqref{eqn.GRHC}$ _2 $ has the same structure 
as the non-relativistic 
SHTC 
heat conduction equation~\cite{SHTC-GENERIC-CMAT,Rom1998}.

\subsubsection{The symmetry of the thermal stress}

Let us now consider the question of symmetry of the energy-momentum. Thus, 
in new terms, the thermal stress can be written as
\begin{equation}\label{therm.stress.sym}
\pot_\mu E_{\pot_\nu} = -(\potw_\mu +\Temp u_\mu)(\entropy u^\nu + 
\E_{\potw_\nu}) =
-( \potw_\mu \E_{\potw_\nu} + \Temp \entropy u_\mu u^\nu + \entropy \potw_\mu 
u^\nu + \Temp 
\E_{\potw_\nu} 
u_\mu)
\end{equation}
which is symmetric if and only if 
\begin{equation}\label{symm.cond}
\entropy \potw^\nu = \Temp \E_{\potw_\nu}.
\end{equation}
This, in fact, implies that 
$ \ETJ_\Temp = \ETW_\Temp + 2 \Temp \ETW_\potw = \entropy - 2 \Temp \E_\potw = 
0$, that 
is $ 
\ETJ(\Temp,J) = 
\ETJ(J)$ does not depend explicitly on $ \Temp $ and we can identify $ E \equiv 
\ETJ $ in \eqref{energies}.

\subsubsection{Heat flux}
Recall that, in the SHTC theory \cite{Rom1998,SHTC-GENERIC-CMAT,DPRZ2016}, the 
heat flux 
is 
introduced as
\begin{equation}\label{heat.flux}
q^\mu := \E_\entropy \E_{\potw_\mu} = \Temp \E_{\potw_\mu}
\end{equation}
hence, if condition \eqref{symm.cond} holds, the thermal stress 
\eqref{therm.stress.sym} can be equivalently 
rewritten as
\begin{equation}\label{therm.stress.sym2}
\pot^\mu E_{\pot_\nu} = 
-(\entropy \E_\entropy u^\mu u^\nu  + \potw^\mu \E_{\potw_\nu} + q^\nu u^\mu + 
u^\nu 
q^\mu),
\end{equation}
which is manifestly symmetric. Hence, the total energy-momentum for a 
relativistic heat-conducting fluid (anisotropic shear stress is omitted) reads
\begin{equation}
-\EMT{\nu}{\mu} =
(\entropy \E_\entropy -\E) u^\nu u_\mu + (-\rho \E_\rho - \entropy \E_\entropy 
+ \E)h^\nu_{\ \mu} -
\entropy \E_\entropy u^\nu u_\mu -\potw_\mu \E_{\potw_\nu} - q^\nu u_\mu - 
u^\nu q_\mu,
\end{equation}
so that we have
\begin{equation}
\EMT{\nu}{\mu} =
\E u^\nu u_\mu + p\, h^\nu_{\ \mu}
+ \potw_\mu \E_{\potw_\nu} + q^\nu u_\mu + u^\nu q_\mu,
\end{equation}
with $ p := \rho \E_\rho + \entropy \E_\entropy - \E $ being the pressure. 
Here, 
the last 
two terms are conventional \cite{Van2008,RezzollaZanottiBook}, while the term $ 
\potw_\mu 
\E_{\potw_\nu} $ is due to the 
non-equilibrium nature of the theory. The presence of such a term can be 
in particular justified 
based on macroscopic equations for non-equilibrium gas 
flows (moments equations) \cite{Torrilhon2016} derived from the Boltzmann 
kinetic equation. Such a term also 
appears in the GENERIC formulation of relativistic heat conduction 
\cite{Ottinger1998,Ottinger2019} which we found to be equivalent to the SHTC  
variational formulation. However, in 
the near-equilibrium settings (diffusive regime), this term can be ignored due 
to its smallness, see Sec.\,\ref{sec.asymptotic}.

\subsection{Hyperbolicity} 
\label{sec.therm.consistency}

Hyperbolic PDEs provide a natural framework for modeling time-dependent 
physical phenomena at a macroscale because they have a locally well-posed initial value 
problem and admit only finite speeds for perturbation 
propagation usually associated with the sound speeds in matter, which are subluminal 
(causality). 
Moreover, local well-posedness is also a critical property of a PDE system for its 
consistent numerical resolution. In particular, it is one of the goals of this paper to 
obtain a 
relativistic version of our hyperbolic equations for viscous momentum  
\cite{HPR2016,DPRZ2016} and heat transfer 
\cite{Rom1998,DPRZ2016,SHTC-GENERIC-CMAT}. However, because 
hyperbolicity of a 
non-homogeneous first-order system is defined by only its leading order 
terms, we can consider this 
question even before the introduction of the irreversible source terms (low-order terms) 
in 
Sec.\,\ref{sec.full.model}.

 From the mathematical standpoint, 
hyperbolicity of a first-order
non-linear PDE system in a domain $ \Omega $ is equivalent to its 
\emph{strong linear stability} at 
any 
point of $ \Omega $ \cite{Despres2017}. 
Recall that the \emph{strong} linear stability implies not only that the 
characteristic 
velocities of the considered PDE system (eigenvalues) are \emph{real} but also that a full set of 
eigenvectors exists, that is the quasilinear form of the PDE system is diagonalizable. In turn, 
this two 
conditions results in that the norm of Fourier-Laplace modes (the 
solution of the linearised PDE system) are \textit{uniformly bounded} in time.
Therefore, it is in principle equivalent to study 
hyperbolicity of a model in Eulerian and Lagrangian frames (if only the two 
have the same number and types of unknowns\footnote{This is not always the case. For example, the 
	Eulerian formulation of the Euler equations of ideal fluids requires 5 fields (density, 
	momentum 
	and energy), while its Lagrangian formulation requires extra nine fields for the deformation 
	gradient, e.g. see \cite{Despres2017}.}) because their Fourier-Laplace modes  are connected by 
	the 
solution-dependent non-singular transformation \eqref{euler.eq.change.J} and thus, the growth type 
of the 
Fourier-Laplace mode cannot 
qualitatively 
change, i.e. from a uniformly bounded to unbounded behavior (blow-up solution) and 
\textit{vice versa}.

Having said that, it has appeared that it is much easier to prove hyperbolicity, and 
even 
\textit{symmetric hyperbolicity}, of the Lagrangian system \eqref{eqn.motion.Lag} and 
\eqref{heat.EL} while hyperbolicity of its Eulerian counterpart that is summarized 
later in 
\eqref{GRGPR} and used in the numerical simulation in Sec.\ref{sec.numerics} follows from the 
discussion above. In addition to the reasoning discussed at the beginning of 
Sec.\ref{sec.reversible}, this is another reason why we employ the Lagrangian frame in our 
theoretical considerations. Also, recall that for the first-order symmetric 
hyperbolic equations, 
a non-increasing energy 
functional can be easily obtained, see e.g. \cite{Rom1986heat,Godunov1973,Hiscock1983}, which can 
be used to 
obtain estimates for the norm of the solution and its derivatives.


Lagrangian governing equations 
\eqref{eqn.motion.Lag} and \eqref{heat.EL}, 
after introducing new variables $ m_\mu :=\Lag_{u^\mu} $, $ S :=\Lag_{T} $, and 
a new potential $ \scrU := u^\mu \Lag_{u^\mu} + T \Lag_{T} - \Lag $ as Legendre 
conjugates,
read as
\vspace{-0.25cm}
\begin{subequations}\label{PDE.Lagr}
	\begin{align}
	\pd_\tau\,m_\mu - \pd_\sA\,\scrU_{\F{\mu}{\sA}} &= 0,
	\qquad
	\pd_\tau\, \F{\mu}{\sA} - \pd_\sA\, \scrU_{m_\mu} = 0, \\
	\pd_\tau S + \pd_\sA \scrU_{\pot_\sA} &= 0, 
	\qquad 
	\partial_\tau \pot_\sA + \partial_\sA \scrU_{S} = 0,
	\end{align}
	where $ \pd_\tau = \pd_0 $ is the Lagrangian time derivative and $ \tau $ is the proper time.
	This is exactly the system of conservation laws studied in 
	\cite{Godunov1996,SHTC-GENERIC-CMAT} and which is symmetrizable and 
	compatible with the first law of thermodynamics (thermodynamical 
	compatibility), i.e. it 
	admits an extra conservation law for the potential $ \scrU $:
	\begin{equation}\label{extra.law}
	\pd_\tau \scrU - \pd_\sA(\scrU_{m_\mu} \scrU_{\cgrad{\sA}{\mu}} - 
	\scrU_\Entropy \scrU_{\pot_\sA}) = 0,
	\end{equation}
\end{subequations}
which can be interpreted as the total energy conservation. Moreover, this system is \emph{ 
symmetric 
hyperbolic} \cite{Friedrichs1958} if 
the potential $ \scrU $ is convex, see details  
in \cite{SHTC-GENERIC-CMAT}, that is its characteristic velocities (they give the 
perturbation propagation velocities) are guarantied to be always real and 
the quasilinear form of \eqref{PDE.Lagr} is diagonalizable.
	Moreover, the 
potential $ \scrU $ is designed in such a way 
that the characteristic velocities coincide with the high-frequency limit of the sound speeds of 
the material 
under consideration, e.g. see \eqref{sum.E}, which can be taken from experimental data 
\cite{Greenspan1956,DPRZ2016,HYP2016} or estimated from microscopic theories. Of 
course in reality, these 
sound speeds are always less than the speed of light and hence, the model is causal. Thus, similar 
to the 
M\"uller-Israel-Stewart theory 
\cite{RezzollaZanottiBook}, our theory has the longitudinal, transversal or shear, and thermal 
sound velocities, e.g. see \cite{DPRZ2016}.

\section{Relativistic heat-conducting viscous media} 
\label{sec.full.model}

In this section, we finalize the formulation of the SHTC equations 
for relativistic heat-conducting viscous/elastoplastic media by 
specifying the irreversible part of the time evolution. Note that the 
differentiation between viscous and elastoplastic media is achieved via a 
proper choice of the dependence of the relaxation time $ \taus $, see 
\eqref{thetas.thetah}, on the 
state variables, e.g. see \cite{Hyper-Hypo2019,Busto2019}, and does not 
depend on 
the reversible part of the time evolution. We then
discuss 
thermodynamic consistency of the governing equations, close the system by 
providing an example of 
the energy potential, and recover \emph{effective} shear viscosity and heat 
conductivity of our 
theory.

The system of SHTC governing equations for general relativistic heat-conducting 
viscous/elastoplastic media reads as follows
\begin{subequations}\label{GRGPR}
	\begin{align}
	&\nabla_\nu(\E u^\nu u_\mu + p\, h^\nu_{\ \mu} + \dist{a}{\mu} 
	\E_{\dist{a}{\nu}}
	+ \potw_\mu \E_{\potw_\nu} + q^\nu u_\mu + u^\nu q_\mu	) 
	= 0, \label{GRGPR.T}\\
	&u^\nu (\nabla_\nu\dist{a}{\mu} - \nabla_\mu\dist{a}{\nu}) = 
	-\frac{1}{\thetas} 
	\GeffContr{ab} g_{\mu\nu} \E_{\dist{b}{\nu}},\label{GRGPR.A}\\
	&	\nabla_\mu( \entropy u^\mu + \E_{\potw_\mu}) = \frac{1}{\Temp}\left( 
	\frac{1}{\thetas} \GeffContr{ab} g_{\mu\nu} \E_{\dist{a}{\mu}}\E_{\dist{b}{\nu}} +
	\frac{1}{\thetah}g_{\mu\nu} \E_{\potw_\nu}\E_{\potw_\mu}
		\right) \geq 0,\label{GRGPR.s}\\
	&u^\lambda(\nabla_\lambda \potw_\mu - \nabla_\mu \potw_\lambda) + 
	h^\lambda_{\ \mu} Q_\lambda = 
	-\frac{1}{\thetah}g_{\mu\nu} \E_{\potw_\mu},\label{GRGPR.J}\\
	& \nabla_\mu(\rho u^\mu) = 0,\label{GRGPR.rho}
	\end{align}
where 
	\begin{equation}
		p = \rho \E_\rho + s \E_s - \E, 
		\qquad
		T = \E_s, 
		\qquad %
		q^\mu = T \E_{\potw_\mu},
		\qquad %
		Q_\mu = \nabla_\mu T + T u^\lambda \nabla_\lambda u_\mu,
%
	\end{equation}
\end{subequations}
The left hand-side in \eqref{GRGPR} represents the reversible part of the time 
evolution derived in Sec.\ref{sec.Var.Lagr} and \ref{sec.Heat}, while the newly 
added relaxation source terms on the right 
hand-side of \eqref{GRGPR} represent the irreversible part and can be viewed as 
gradients of the quadratic dissipation potential \cite{SHTC-GENERIC-CMAT}. The 
total energy $ \E $ is 
left unspecified so far. However, one may clearly notice that the definition of 
the reversible and irreversible terms depends on 
the specification of the total energy $ \E $ and will be provided in  
Sec.\,\ref{sec.EOS} as well as the specification of the relaxation parameters $ 
\thetas $ and $ \thetah $.

One of the non-trivial differences between the reversible equations (with zeros 
on the right hand-side) and irreversible ones is in the substitution of the 
configuration gradient $ \cgrad{a}{\mu} $, which can be seen as a holonomic 
basis tetrad, by the distortion field $ 
\dist{a}{\mu} $, which is a non-holonomic basis tetrad, i.e. it is not a  
gradient of the mapping \eqref{eqn.motion.laws1}. This is discussed in detail in 
\paperI\ in the relativistic setting and in 
\cite{PRD-Torsion2019,HPR2016,DPRZ2016} in the non-relativistic 
setting. It is implied though that $ \detA = \detxi $.

One may clearly see that the GR version of SHTC equations 
\eqref{GRGPR} shares 
a lot of common structural features with the non-relativistic SHTC equations, 
e.g. 
\cite{SHTC-GENERIC-CMAT}. We, therefore, expect that the relativistic 
equations \eqref{GRGPR} also 
admit a Hamiltonian formulation similar to \cite{Ottinger1998,Ottinger-book}, that is the 
reversible part of the time evolution is generated by the anti-symmetric Poisson brackets, while 
the irreversible part can be generated by a symmetric bracket, see \cite{Ottinger-book} and Sec.4 
in \cite{PKG-Book2018}. 
However, to show 
this, one may need to obtain a genuinely 4-dimensional formulation of the 
GENERIC approach which 
currently relies on the explicit treatment of the time coordinate and therefore, is not truly 
4-dimensional.
Nevertheless, one may note that the irreversible part of the GR SHTC 
equations \eqref{GRGPR} is 
identical to its non-relativistic counterpart in \cite{SHTC-GENERIC-CMAT} and 
thus, it can be seen 
as the gradient of the quadratic dissipation potential , see 
\cite{SHTC-GENERIC-CMAT}, and hence, 
it 
can also be viewed as it is generated by a symmetric dissipative bracket as in 
the branch of GENERIC 
developed in \cite{Ottinger1998,Ottinger-book}.

\subsection{Linear stability}
One of the fundamental observations of the non-equilibrium thermodynamics is that an 
out-of-equilibrium system, if left free of external stimuli, tends towards the global equilibrium 
state \cite{PKG-Book2018}. From the mathematical standpoint, this is equivalent to that the 
equilibrium solution of \eqref{GRGPR} is linearly stable. Therefore, introduction of the 
dissipative 
relaxation terms in \eqref{GRGPR} should not contradict to this observation. To prove that the 
out-of-equilibrium solutions of \eqref{GRGPR} tend towards the equilibrium state, or that the 
perturbations of the equilibrium state decay in time, is again easy to 
do in the Lagrangian frame, that is for system \eqref{PDE.Lagr} supplemented with the same type
source terms as in \eqref{GRGPR} generated by the quadratic (convex) dissipation potential (see the 
discussion just before this subsection). Indeed, 
the 
resulting Lagrangian symmetric hyperbolic relaxation system, e.g. 
see \cite{SHTC-GENERIC-CMAT}, 
fulfills the conditions (symmetrizability and the non-negative definiteness of the dissipation 
matrix) 
of the class of first-order hyperbolic PDE systems studied by Yong and co-authors in 
\cite{Yong1999,Yong2015} for 
which it is proven that the equilibrium state is linearly stable.

\subsection{Thermodynamic consistency}

The choice of the entropy production source term in eq.\eqref{GRGPR.s} is 
guided by both laws of 
thermodynamics, i.e. it has to be non-negative in order to guarantee the 
non-decreasing of the physical entropy in irreversible process (second law of 
thermodynamics), but it also allows 
to conserve the total energy (first law of thermodynamics). Indeed, system \eqref{GRGPR} is an overdetermined 
PDE system, that is there are more equations than unknowns. Hence, if all 
equations 
are compatible with each other, one of the equations should be the consequence 
of the others. 
Thus, one can check that the 0-\textit{th} equation in \eqref{GRGPR.T}, the 
total energy conservation law, is a  
linear combination of the remaining equations\vspace{-0.45cm}
\begin{multline}\label{summation}
\nabla_\nu(\E u^\nu u_0 + p\, h^\nu_{\ 0} + P^\nu_{\ 0}	) = -\frac{u^i}{u^0} 
\nabla_\nu(\E u^\nu u_i + p\, h^\nu_{\ i} + P^\nu_{\ i}	) - 
\frac{\E_{\dist{a}{\mu}}}{u^0}u^\nu (\nabla_\nu\dist{a}{\mu} - 
\nabla_\mu\dist{a}{\nu}) \\
-\frac{\E_s}{u^0} \nabla_\mu( \entropy u^\mu + \E_{\potw_\mu})
-\frac{\E_{\potw_\mu}}{u^0}(u^\lambda(\nabla_\lambda \potw_\mu - \nabla_\mu 
\potw_\lambda) + h^\lambda_{\ \mu} Q_\lambda) - \frac{\E_\rho}{u^0}  
\nabla_\mu(\rho u^\mu),
\end{multline}
where $ P^\nu_{\ \mu} = \dist{a}{\mu} \E_{\dist{a}{\nu}} + \potw_\mu 
\E_{\potw_\nu} + q^\nu u_\mu + u^\nu q_\mu $. The source terms in \eqref{GRGPR} 
are designed in such a way that they canceled out during the summation 
\eqref{summation}, that is the total energy is indeed conserved.

Finally, we note that the rest mass conservation law~\eqref{GRGPR.rho} is, in 
fact, not an independent equation, but can be derived from the configuration 
gradient \eqref{eqn.motion.Eul1}$ _2 $ or distortion evolution equation 
\eqref{GRGPR.A}, e.g. see \paperI. Nevertheless, it is convenient to formally 
treat the 
rest mass 
density $ \rho := \rho_0 \detA/\sqrt{-g} $ as an independent variable \paperI, 
which should be taken into account in the derivatives $ \E_{\dist{a}{\mu}} $ 
and $ \E_\rho $. Here, $ \rho_0 $ is the reference mass density.
\vspace{-0.15cm}


\subsection{Closure: equation of state}\label{sec.EOS}

As we have seen, the SHTC system of governing equations is formulated without specifying 
the Lagrangian  density $ \LED $ or energy $ \E $. We say that the energy potential $ \E 
$ is the generating potential for system \eqref{GRGPR} which has to be 
specified in order 
to close the system.

Because $ \E $ has to be a proper scalar, it has to depend on the the tensor 
fields via their invariants, 
including mixed invariants of two or more fields. The subsystems of 
\eqref{GRGPR} can be, 
therefore, non-linearly coupled via a specific closure, i.e. a specific choice 
of the potential $ \E $. 
In this paper, however, we give an example of a rather simple closure (quadratic energy) which 
is discussed in the following subsection.

We assume the following decomposition of the total energy potential $ \E $
\begin{equation}
	\E(\rho,\entropy,w_\mu,\dist{\sA}{\mu}) = \rho 
	(1 + \varepsilon(\rho,\Entropy,w_\mu,\dist{\sA}{\mu})), 
\end{equation}
where $ \Entropy = s/\rho $ is the specific entropy (per unit of rest mass).


Furthermore, we shall need the material metric $ \Geff{\mu\nu} $, see \paperI,
\begin{equation}\label{eqn.Geff}
\Geff{\mu\nu} := \LagrProjEff_{ab} \dist{a}{\mu}\dist{b}{\nu} = 
\Kab{\sA\sB}\dist{\sA}{\mu}\dist{\sB}{\nu},
\qquad
\LagrProjEff_{ab} := \Kab{ab} + \uLag_a \uLag_b =
\left(\begin{array}{rccc}
0 & 0 & 0 & 0\\ 
0 &   &   &  \\ 
0 &   & \Kab{\sA\sB} & \\
0 &   &   &  
\end{array}\right),
\end{equation}
where we have defined the Lagrangian matter projector $ \LagrProjEff_{ab} $
in the local relaxed frame, and $ \uLag^a = \frac{\pd \myxi^a}{\pd \proptime} = 
(1,0,0,0) $, $ 
\uLag_a = \Kab{ab}\uLag^b = 
(-1,0,0,0)$ is the Lagrangian 4-velocity. Following our papers on Newtonian continuum 
mechanics~\cite{DPRZ2016,DPRZ2017}, we shall decompose the material 
metric $ \Geff{\mu\nu} $ into a traceless part $ \devG{\mu\nu} $ 
and a spherical part:
\begin{equation}\label{eqn.devG}
\Geff{\mu\nu}  = \devG{\mu\nu} + 
\frac{\Geffmix{\lambda}{\lambda}}{3}
h_{\mu\nu}, \quad \text{ where } \quad
\devG{\mu\nu} := \Geff{\mu\nu} - 
\frac{\Geffmix{\lambda}{\lambda}}{3}
h_{\mu\nu}.
\end{equation}
Note that, in this definition, $ \devG{\mu\nu} $ refers to the spatial 
projector $ h_{\mu\nu} $ and not to the full 
spacetime metric $ g_{\mu\nu} $. We then use the norm of the traceless part (here one has to use 
that $ 
h^{\mu}_{\ \mu} = g^{\mu\lambda} 
h_{\lambda\mu} = 3 $)
\begin{equation}\label{eqn.dev.norm}
\devGmix{\lambda}{\nu}\devGmix{\nu}{\lambda} = I_2 - 
I_1^2/3,
\qquad
I_1 = \Geffmix{\mu}{\mu}, 
\qquad 
I_2 = \Geffmix{\mu}{\nu}\Geffmix{\nu}{\mu},
\end{equation}
as an indication of the presence of non-volumetric (tangential) deformations, 
and define the 
specific energy $ \varepsilon(\rho,\Entropy,\potw_\mu, \dist{\sA}{\mu}) 
$\vspace{-0.35cm}
\begin{equation}\label{sum.E}
	\varepsilon(\rho,\Entropy,\dist{\sM }{\mu},\potw^\mu) = \varepsilon^{\textrm 
	eq}(\rho,\Entropy) + 
	\frac{\cs^2}{4} \devGmix{\lambda}{\nu}\devGmix{\nu}{\lambda} + 
	\frac{\alphah^2}{2} \potw^\mu \potw_\mu,
\end{equation}
where $ \varepsilon^{\textrm eq} $ is given by a hydrodynamic equation of state (EOS), which can be 
either the ideal or stiffened gas EOS (in the case of liquids or solids) or a general tabulated 
one. Here, $\cs$ 
denotes the sound speed of propagation of shear perturbation, that is the 
characteristic velocity of propagation of shear perturbation in the most 
non-equilibrium state, i.e. when the associated relaxation parameter $ \taus $ 
goes 
to infinity, $ \taus \to \infty $.  The parameter $ \alphah $ is related to 
the characteristic velocity of propagation of thermal perturbations $ \ch $ 
(so-called second sound) in the non-equilibrium state ($ \tauh \to 
\infty $) as $\ch^2 = \alphah^2 T / c_\sV$, where $ \tauh $ is the thermal 
relaxation time scale, and  $ c_\sV $ is the specific 
heat capacity at constant volume \cite{DPRZ2016}. For such a 
specification of $ \E $, we have
\begin{subequations}
	\begin{gather}
		\E_{\dist{\sA }{\mu}} =
		\rho\, \cs^2
		\KMN_{\sA \sB}\dist{\sB}{\lambda}g^{\lambda\alpha} \devG{\alpha\beta} 
		h^{\beta 
		\mu}, 
		\qquad
		\dist{\sA }{\mu}  \E_{\dist{\sA }{\nu}} = \rho\, \cs^{2} 
		\devGmix{\nu}{\lambda} 
		\Geffmix{\lambda}{\mu}, \label{eqn.shear.stress}\\
		\potw_\mu \E_{\potw_\nu} = \rho \alphah^2 \potw^\nu \potw_\mu,
		\qquad
		q^\mu = T \rho \alphah^2 \potw^\mu, 
		\qquad
		T = \E_\entropy =\varepsilon^{\textrm eq}_\Entropy
		\label{eqn.thermal.stress}
	\end{gather}
\end{subequations}
Based on the dimensional reasoning and asymptotic analysis performed in the 
following section, it is convenient to define the relaxation 
parameters $ \thetas $ and $ \thetah $ as follows 
\cite{GRGPR,DPRZ2016,SHTC-GENERIC-CMAT}
\begin{equation}\label{thetas.thetah}
	\thetas = \rho_0\, \taus \cs^2 \Geffmix{\lambda}{\lambda}/3,
	\qquad
	\thetah = \rho\, \alphah^2 \tauh.
\end{equation}

%

\subsection{Asymptotic analysis}\label{sec.asymptotic}

In this section, via a formal asymptotic analysis performed for the closure 
\eqref{sum.E}, we demonstrate that in the 
asymptotic relaxation limit $ \taus \to 0 $ and $ \tauh \to 0 
$, the leading terms of our equations are identical to the relativistic 
Navier-Stokes-Fourier equations \cite{RezzollaZanottiBook}. Although, the 
latter are 
parabolic and 
non-causal and known to  
be unstable \cite{Hiscock1983,Hiscock1985,Hiscock1988}, it is the high-order 
terms of our theory which should be responsible for the stability of the 
solution.
Via a formal 
asymptotic analysis, we express the effective transport coefficients of the 
theory such as 
shear viscosity and  
heat conductivity in terms of the characteristic velocities $ \cs $, $ \ch $ 
and the relaxation times $ \taus $ and $ \tauh $. 


\paragraph{Effective heat conductivity.} 

Assuming that the thermal impulse can 
be expanded as
\begin{equation}\label{therm.imp.expand}
\potw_\mu = \potexp{0}_\mu + \tauh \potexp{1}_\mu + 
\tauh^2\potexp{2}_\mu+\ldots,
\end{equation}
we plug it in the equation \eqref{w.PDE}, where the irreversible terms 
are now added:
\begin{equation}
u^\lambda\nabla_\lambda(\potexp{0}_\mu + \tauh \potexp{1}_\mu) + 
(\potexp{0}_\lambda + \tauh 
\potexp{1}_\lambda)\nabla_\mu u^\lambda + h^\lambda_{\ \mu} Q_\lambda + \ldots 
= 
-\frac{\rho 
\alphah^2}{\thetah}\potw_{\mu},
\end{equation}
where dots `$ \ldots $' mean higher order terms.
It is convenient to define $ \thetah $ as in \eqref{heat.conductivity}.
Then, collecting terms with equal orders of power of $ \tauh $, we obtain that 
at the zeroth order $ \potexp{0}_\mu = 0 $, while at the first order
$
\potexp{1}_\mu = -h^\lambda_{\ \mu} Q_\lambda.
$
Therefore, by truncating expansion \eqref{therm.imp.expand} to first order,  
the solution can be approximated as $ \potw_\mu = -\tauh h^\lambda_{\ \mu} 
Q_\lambda $. Then, 
using that the heat flux 4-vector is defined 
as $ q^\mu = \Temp \E_{\potw_\mu} $ in 
\eqref{heat.flux}, one has
\begin{equation}
q^\mu = \Temp \E_{\potw_\mu} = \Temp \rho \alphah^2 \potw^\mu = - \Temp \rho 
\alphah^2 g^{\mu\nu}\tauh h^\lambda_{\ \nu} Q_\lambda,
\end{equation}
that is the effective heat conductivity $ \kappa^{\textrm eff} $ of our model in 
the diffusive regime ($ 
\tauh \ll 1 $) can be defined as
\begin{equation}\label{heat.conductivity}
\kappa^{\textrm eff} = \rho\, T \tauh \alphah^2 = \rho\, c_{\sV} \tau \ch^2.
\end{equation}
\paragraph{Effective shear viscosity.} In \paperI, it was shown 
that, 
in the absence 
of heat conduction ($\ch=0$), a formal asymptotic 
expansion reveals the structure of the leading terms of viscous stress 
$\Cauchymix{\nu}{\mu} = \dist{\sA}{\mu}\E_{\dist{\sA}{\nu}}$  in 
the 
asymptotic 
relaxation limit when $\taus \to 0$,  
\begin{equation}\label{eqn.NSE.stress}
\Cauchy{\mu\nu} = - \frac{1}{6} \rho_0 \taus \cs^2 \left( 
h_{\mu\lambda}\nabla_\nu u^\lambda
+ h_{\lambda\nu}\nabla_\mu u^\lambda - 
\frac{2}{3} (h^\alpha_{\ \lambda} \nabla_\alpha u^\lambda) h_{\mu\nu} + 
{ u^\lambda \nabla_\lambda h_{\mu\nu}}
\right),
\end{equation}
which is equivalent to the Landau-Lifshitz version of the relativistic Navier-Stokes stress 
\cite{Landau-Lifshitz6,RezzollaZanottiBook} with the effective viscosity $\mu = 
\frac{1}{6} \rho_0 \taus \cs^2$. 

\section{Numerical validation}\label{sec.numerics}

\subsection{3+1 formulation}\label{sec.split}

In a few words, in the $3+1$ foliation of spacetime it is a usual procedure to 
project the $ 4d $ covariant governing equations into a set of $ 3d$ equations 
by 
using the so called \emph{spatial} and \emph{temporal} projection operators,  
$\gam{\mu\nu}$ and $N_{\mu\nu}$, respectively, such that $\g_{\mu\nu} = 
\gam{\mu\nu} + N_{\mu\nu}$, e.g. see \cite{RezzollaZanottiBook}. 
In this way, any symmetric rank-2 tensor can be decomposed in its spatial 
and temporal components, e.g.
\begin{align}
T^{\mu\nu} & = S^{\mu\nu} + S^\mu n^\nu + n^\mu S^\nu + \mathcal{U} n^\mu n^\nu 
\label{eq:divT},
\end{align}
where, 
\begin{equation}
	S^{\mu\nu} :=\gammix{\mu}{\alpha} \gammix{\nu}{\beta} T^{\alpha \beta}, 
	\qquad  
	S^\mu:=-\gammix{\mu}{\alpha} n_\beta T^{\alpha \beta}, 
	\qquad 
	\mathcal{U}:=n_\alpha 
	n_\beta T^{\alpha \beta} \label{eq:SSU}  
\end{equation}
In particular, such operators are defined after specifying on the spacetime 
$\mathcal{V}^4$ a proper field of local (Eulerian) observers travelling with 
non-constant 4-velocity $n^{\mu}$. In particular, one can show that 
\cite{RezzollaZanottiBook}
\begin{gather*}
g_{\mu\nu} = \left( \begin{array}{cc} -\lapse^2 + \shift_i\shift^i & \shift_i 
\\ \shift_i & \gamma_{ij} \end{array}\right),
\qquad
g^{\mu\nu} = \left( \begin{array}{cc} -1/\lapse^2   & \shift^i / \lapse^2  \\   
\shift^i  /  \lapse^2   & \gamma^{ij} - \shift^i\shift^j / \lapse^2 
\end{array}\right) ,\\
n_{\mu} = -\lapse \nabla_{\mu} t = (-\alpha, 0_i) \,,
\quad
n^{\mu} = (  1 /\lapse , - \shift^i / \lapse ) \,,\quad  n_\mu n^\mu=-1,\\
\gamma_{\mu\nu}:= g_{\mu\nu} + n_\mu n_{\nu}\,, \;\;
\qquad 
N_{\mu\nu}:= -n_\mu n_\nu,
\end{gather*}
where $t$ is chosen to be the  time coordinate, $\lapse$ is the \textit{lapse 
function}, $\shift_j$ is the \textit{shift vector}, and $\gamma_{ij}$ are 
the spatial components of the spatial metric $ \gam{\mu\nu} $. The 
corresponding  identities related to the medium 4-velocity $u^\mu$ are
\begin{gather*}
u^\mu = \Lor \left( n^\mu + v^\mu\right)\,, 
\quad
\Lor := - n_\mu u^\mu = \lapse u^t = (1-v_iv^i)^{-1/2} = (1-v^2)^{-1/2} \,,\\
\boldsymbol{\gamma}\cdot  \boldsymbol{u}
= ( \delta^{\mu}_{\ \nu} + n^\mu n_{\nu} ) u^\nu =  \Lor v^\mu\,,
\quad
v^i  =  u^i/\Lor + \shift^i/\lapse\,,
\end{gather*}
where $ \Lor $ is the Lorentz factor.

Then, after denoting with $\boldsymbol{V}$ the array of the so-called 
33~\emph{primitive variables} 
\begin{align}
\boldsymbol{V} := \left( \rho, v_j, p, \dist{i}{j}, J_j, \KAB_{\sA\sB}, \lapse, 
\shift^j, \gamma_{ij} \right), 
\end{align}
%
where $\KAB_{\sA\sB}$ are the material components of the matter-time metric $ 
\Kab{ab} $.
The so-called 33~conserved variables $\boldsymbol{Q}(\boldsymbol{V})$ can be 
easily
expressed in terms of the primitive variables via the relations
\begin{align}
& D := \rho \Lor\,,\qquad \boldsymbol{S} := \rho h \Lor^2 
\boldsymbol{v},\qquad  \mathcal{U} := \rho h \Lor^2 - p \,.
\end{align}
Here, $\mathcal{U}$ is the conserved energy density,
$h=1+\epsilon+p/\rho$ is the specific enthalpy and $\epsilon$ is the
specific internal energy \cite{RezzollaZanottiBook}.
Then, the chosen state vector $\boldsymbol{Q}$ of {conserved}
variables with respect to the PDE system (\ref{eq:finalPDE}) is defined as
\begin{align}
\boldsymbol{Q} := \left( \sqrt{\gamma}\, D, \sqrt{\gamma}\, S_j, \sqrt{\gamma}\,
\mathcal{U},  \dist{i}{j}, J_j, \KAB_{\sA\sB}, \lapse, \shift^j, \gamma_{ij} 
\right)\,.
\end{align}
While the transformation from primitive to conserved variables is explicit and 
straightforward,
in this work, the inversion of the 
primitive to conserved function is computed iteratively. 

Eventually, after a standard $3+1$ foliation of spacetime 
\cite{Gourgoulhon2012,RezzollaZanottiBook}, system 
(\ref{GRGPR}) projected into the Valencia-type formulation reads as follows: 
\begin{subequations}\label{eq:finalPDE}
	\begin{align}
	& \partial_t \left(  \gamma^{\frac{1}{2}} D \right) +\partial_i \left[  
	\gamma^{\frac{1}{2}}D \left(   \lapse v^i -    \shift^i  \right) \right]  = 
	0 , \\
	&\partial_t \left( \gamma^{\frac{1}{2}}    S_j \right) + \partial_i \left[ 
	\gamma^{\frac{1}{2}} 
	\left( \lapse   S^i_{\ j}-\shift^iS_j \right)\right]   
	-\gamma^{\frac{1}{2}} \left(  \frac{1}{2} 
	\lapse S^{ik}\partial_j \gamma_{ik}  + S_i \partial_j \shift^i   - 
	\mathcal{U} 
	\partial_j \lapse \right) = 
	0   ,  \\
	&  \partial_t \left( \gamma^{\frac{1}{2}} \mathcal{U}  \right)    + 
	\partial_i \left[  
	\gamma^{\frac{1}{2}} \left( \lapse S^i   -  \shift^i   
	\mathcal{U}\right)\right] - 
	\gamma^{\frac{1}{2}}  \left(  \lapse S^{ij}  K_{ij} - S^j    \partial_j ,  
	\lapse \right) =   0 ,  \label{finalPDE.energy} \\
	&\pd_t \dist{i}{j} 
	+  \pd_j  \left( \hat{v}^k \dist{i}{k} \right) + \hat{v}^k \left( \pd_k 
	\dist{i}{j}-   \pd_j   
	\dist{i}{k} \right)  = -\frac{1}{\thetas} \dist{i}{\mu}  
	\devGmix{\mu}{j} ,   \\ 
	& \pd_t J_i + \partial_i \left( \hat v^k J_k + \hat{T}  \right) + \hat{v}^k 
	\left( \partial_k J_i - \partial_i J_k \right)
	=  -\frac{1}{\thetah} J_i,  \\ 
	& \partial_t \KAB_{\sA\sB}  +\hat{v}^k\partial_k \KAB_{\sA\sB}  = 0,  \qquad
	\partial_t \alpha = 0,  \qquad \partial_t \beta^i = 0,  \qquad \partial_t 
	\gamma_{ij} = 0,   
	\end{align} 
\end{subequations}
where $ \hat{T} = \lapse T / \Lor $.
%
%
In more detail, the components $S^i_{\ j}$, $S_i$ and $\mathcal{U}$ are defined as 
the 
space-time decomposition (\ref{eq:SSU}) of the energy-stress tensor 
$\EMT{\nu}{\mu}$ that appears in equation (\ref{GRGPR.T}).
Also, we have introduced the definition of the so called \emph{transport 
velocity} as $\hat{v}^i := (v^i + \shift^i)/\lapse$. 
Finally, we stress that due to the fact that system \eqref{GRGPR} is 
overdetermined, see \eqref{summation}, we are free to chose whether the total 
energy or entropy equation is discretized (we chose to discretize the energy 
equation). The remaining quantity is computed from the equation of state. Finally, for the static 
spacetimes, the Cowling approximation \cite{Gourgoulhon2012} 
\begin{align}
\lapse S^{ij}  K_{ij} \equiv \frac{1}{2} S^{ik} \shift^j \partial_j \gamma_{ik} + S^j_{\ i} 
\partial_j\shift^i
\end{align}
can be used for the extrinsic curvature $ K_{ij} $ \cite{RezzollaZanottiBook} in 
\eqref{finalPDE.energy}.

\subsection{Numerical examples}

\paragraph{Heat conduction.} Here we solve a simple Riemann problem that 
involves heat conduction. The initial condition consists in a density jump from 
$\rho_L = 1$ to $\rho_R = 0.9$ located at $x=0.5$. The other state variables 
are globally constant and are chosen as $\bm{v}=\bm{J}=0$, 
$\bm{A}=\bm{I}$ and $p=1$. We furthermore assume a flat Minkowski 
spacetime, hence the lapse function is $\alpha=1$, the shift vector is 
$\beta^i=0$ and the spatial metric tensor is $\gamma_{ij} = \delta_{ij}$. The 
remaining parameters of the model are the ideal gas EOS for $ \varepsilon^{\textrm 
eq} $ in \eqref{sum.E} with $\gamma = \frac{5}{3}$ (the ratio of the 
specific heats), $\cs = 0$, 
$\alphah = 0.8$, $c_\sV=1$ and two different relaxation times $\tauh \ll 1$ are 
considered, namely
 $\tauh=5 \cdot 10^{-3}$ and $\tauh=5 \cdot 10^{-4}$. The chosen initial data lead to a temperature jump of $T_L = \frac{1}{\gamma-1} \frac{1}{\rho_L}$ to $T_R=\frac{1}{\gamma-1} \frac{1}{\rho_R}$. Simulations are run on the domain $\Omega=[0,1] \times [0,1]$ until $t=0.5$ using a fourth order ADER-DG scheme \cite{ADERGRMHD,DPRZ2016,FO-CCZ4} with polynomial approximation degree $N=3$ and $100 \times 4$ spatial elements. A fine grid reference solution is computed with a third order ADER-WENO finite volume scheme \cite{DumbserEnauxToro,DPRZ2016,FO-CCZ4} on 1000 cells. 
 
\begin{figure}[!htbp]
	\begin{center}
		\begin{tabular}{cc} 
			\includegraphics[draft=false,width=0.49\textwidth]{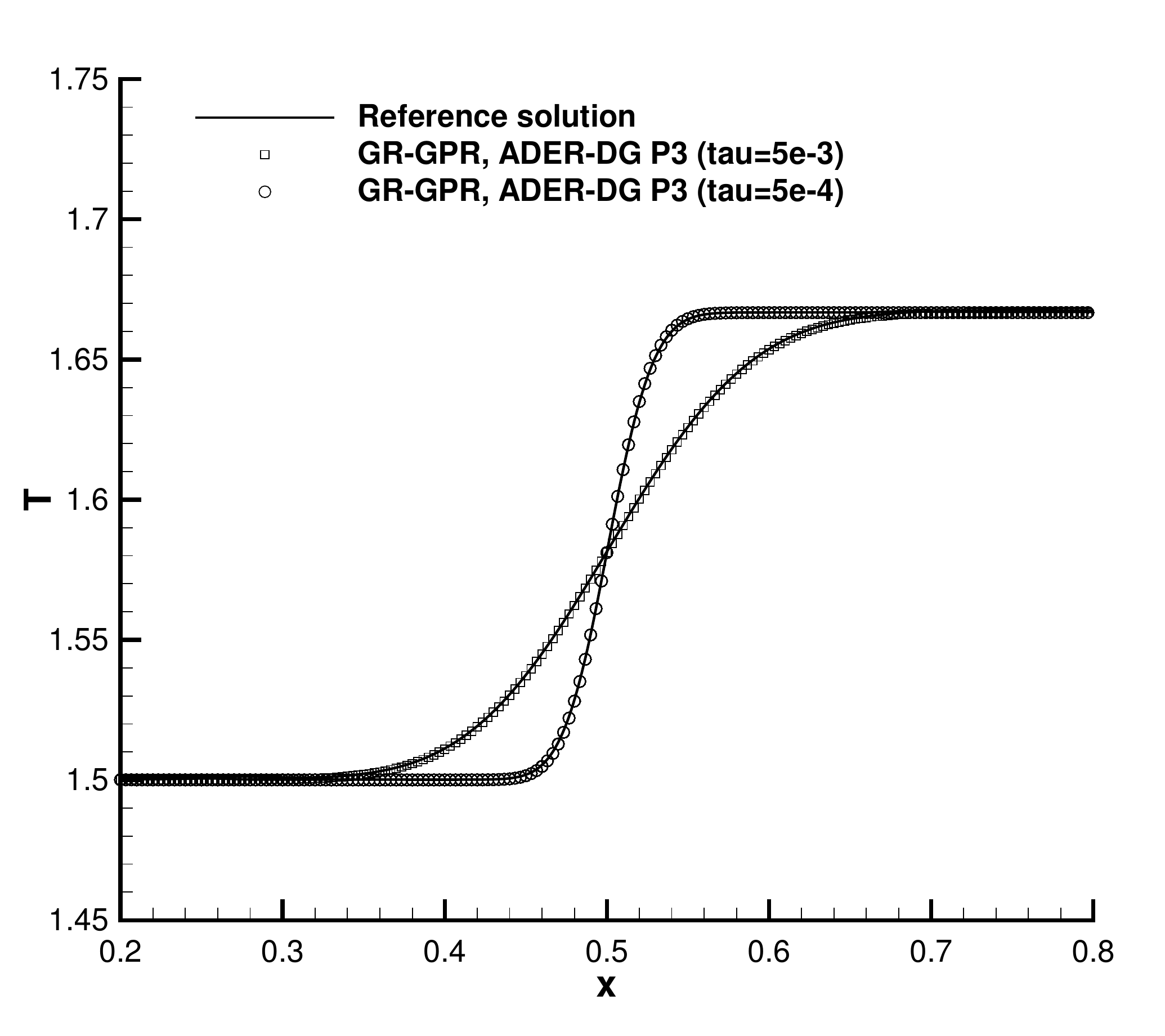}
			  & 
			\includegraphics[draft=false,width=0.49\textwidth]{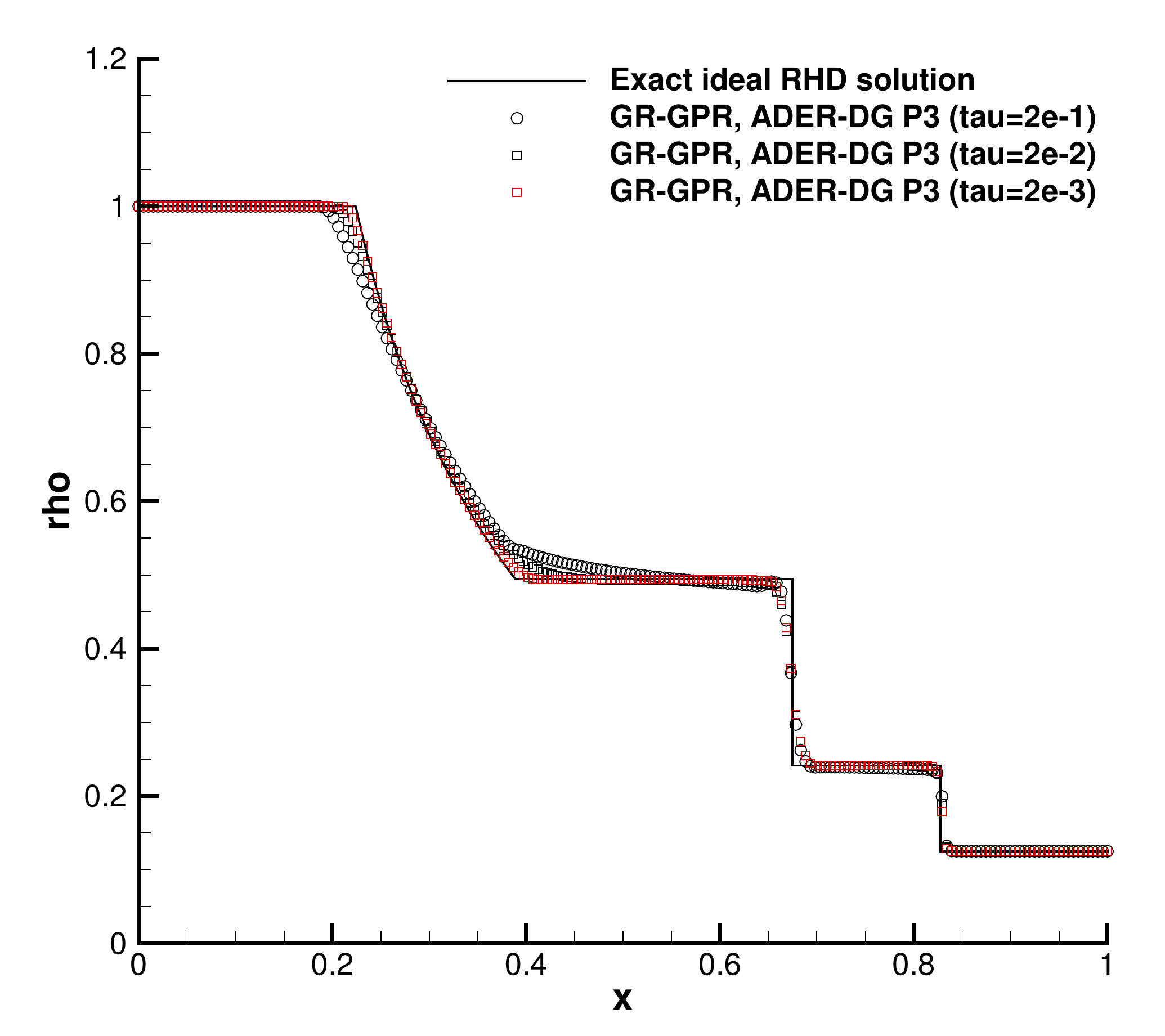}	

		\end{tabular} 
			\caption{Left panel: Heat conduction based on the SHTC model. 
			Temperature distribution at $t=0.5$ for two different relaxation 
			times $\tauh = 5 \cdot 10^{-3}$ and $\tauh = 5 \cdot 10^{-4}$, 
			starting from a temperature jump initially located at $x=0.5$. 
				Right panel: Relativistic Sod shock tube problem solving the 
				SHTC model \eqref{eq:finalPDE} with viscosity and heat 
				conduction with different 
				relaxation times $\tauh = \taus = 2 \cdot 10^{-3}$, $ 2 \cdot 
				10^{-2}$, and $ 2 \cdot 10^{-1}$. As reference, also the exact 
				solution of the Riemann problem of the ideal relativistic Euler 
				equations (RHD) is shown.}   
		\label{fig.results}		
	\end{center}
\end{figure}

 The obtained results are depicted in the left panel of Fig.~\ref{fig.results}, which show the 
typical behavior of heat conduction in the limit of the Fourier law. The agreement of the numerical 
simulation carried out with the high order discontinuous Galerkin finite element scheme and the finite volume reference solution is excellent. 

\paragraph{Relativistic Sod shock tube.} In this last test problem, we solve a 
relativistic version of the Sod shock tube problem for the complete model 
\eqref{GRGPR}, including viscosity and heat conduction. The initial data 
consist of a 
jump located in $x=0.5$ with $\rho_L = 1$, $\rho_R = 0.125$, $\bm{v}=\bm{J}=0$, 
$\bm{A} = \sqrt[3]{\rho} \, \bm{I}$, $p_L = 1$, $p_R = 0.1$, $\alpha=1$, 
$\beta^i = 0$, $\gamma_{ij}=\delta_{ij}$. The remaining model parameters are 
chosen as the ideal gas EOS for $ \varepsilon^{\textrm eq} $ with $\gamma = 
\frac{5}{3}$, $c_\sV=1$, $\ch=0.01$ and $\cs = 0.5$. Three 
different values of relaxation times $\taus=\tauh \ll 1 $ are used in the 
numerical simulations, 
namely $2 \cdot 10^{-1}$, $2 \cdot 10^{-2}$, and $ 2 \cdot 
10^{-3}$. The computational domain is $\Omega = [0,1] \times [0,1]$ and is 
covered with $100 \times 4$ fourth order ADER-DG elements of polynomial 
approximation degree $N=3$. Simulations are run until a final time of $t=0.4$. 
Since the problem under consideration involves shock waves, we make use of the 
\textit{a posteriori} subcell finite volume limiter presented in 
\cite{Dumbser2014}. The computational results are compared against the exact 
solution of the Riemann problem of the ideal relativistic  hydrodynamics (RHD) 
equations, which was kindly provided by Dr. Zanotti, see 
\cite{RezzollaZanotti}. 
The obtained results are depicted in the right panel of 
Figure\,\ref{fig.results}. For small relaxation times, an excellent agreement 
between the stiff relaxation limit of our model 
and the ideal relativistic Euler equations can be noted. 

\section{Conclusion}

We have presented a general relativistic formulation for viscous/elastoplastic 
heat-conducting continuous medium. Such a formulation is a generalization of 
the  
non-relativistic unified formulation for fluid and solid dynamics advanced 
recently in \cite{HPR2016,DPRZ2016,HYP2016} and relies on the theory of first-order
Symmetric Hyperbolic Thermodynamically Compatible (SHTC) equations 
\cite{SHTC-GENERIC-CMAT,God1961,Godunov1996,Rom1998}.

Both transport processes are considered from a non-equilibrium viewpoint, that 
is no local equilibrium assumptions such as Newton's law of viscosity or 
Fourier law of heat conduction are used. We provide a variational formulation 
for the reversible part of the time evolution which makes the model compatible 
with the Euler-Lagrange structure of the Einstein field equations. The 
irreversible 
part is represented by algebraic (no space and time derivatives) 
relaxation-type source terms and can be viewed as gradients of a dissipation 
potential intimately connected with the entropy, see \cite{SHTC-GENERIC-CMAT}. 
We have observed that our variational formulation of the relativistic heat 
conduction provides equivalent equations to the Hamiltonian formulation by 
\"Ottinger 
\cite{Ottinger2019,Ottinger1998} within the GENERIC approach to non-equilibrium 
thermodynamics. The viscous part which is governed by the distortion evolution 
equation \eqref{GRGPR.A} is different from that presented in 
\cite{Ottinger2019,Ottinger1998}. Nevertheless, at least in the Newtonian 
limit, it also admits a 
Hamiltonian 
formulation as proven in \cite{SHTC-GENERIC-CMAT}.

Via a formal asymptotic analysis, we have recovered the effective transport 
coefficients
of our theory in the near equilibrium regime, see Sec.\,\ref{sec.asymptotic}. 
Finally, we presented a 3+1 split of the governing equations in 
Sec.\,\ref{sec.split} which are then solved using the ADER-DG family of 
high-order 
numerical schemes 
\cite{DumbserEnauxToro,ADERGRMHD,DPRZ2016,FO-CCZ4,Dumbser2018a} 
designed specifically for hyperbolic partial differential equations, see 
Sec.\,\ref{sec.numerics}. We solved 
two one-dimensional Riemann problems in the special relativistic settings in 
order to demonstrate the physical consistency and mathematical regularity of 
the numerical solution.

Further research will concern the obtaining of general relativistic versions of 
the SHTC electrodynamics equations in moving medium 
\cite{DPRZ2017,SHTC-GENERIC-CMAT} with resistivity and of hyperbolic equations 
for 
mass transfer \cite{RomDrikToro2010,Romenski2016}.

\enlargethispage{20pt}



\section*{Acknowledgment}
This research has been supported by the European Union's Horizon 2020 Research and 
Innovation Programme under the project \textit{ExaHyPE}, grant no. 671698 (call FET-HPC-1-2014). 
MD also acknowledges funding from the Italian Ministry of Education, University 
and Research (MIUR) via the Departments of Excellence Initiative 2018--2022  
attributed to DICAM of the University of Trento (grant L. 232/2016). MD has 
also received support from the University of Trento in the frame of the 
Strategic Initiative \textit{Modeling and Simulation}. IP greatly acknowledges 
a financial support by Agence Nationale de la Recherche (FR)
(Grant No. ANR-11-LABX-0040-CIMI) within the program ANR-11-IDEX-0002-02. 
Theoretical results obtained by ER in Secs.\,2 and 3 were partially supported 
by the Russian Science Foundation grant (project 19-77-20004).





\small
\bibliographystyle{unsrtnat}

\bibliography{library}

\end{document}